\documentclass[11pt,a4paper,fleqn]{article}
\usepackage[utf8]{inputenc}
\usepackage{xcolor}

\usepackage{multirow,makecell}
\usepackage{graphicx}
\usepackage{acronym}
\usepackage{ifthen}
\usepackage{substr}
\usepackage[left=2cm, top=2cm, right=2cm, bottom=2cm]{geometry}
\usepackage{array}
\usepackage[figuresright]{rotating}
\usepackage{url}
\usepackage[shortlabels]{enumitem}
\usepackage{epsfig}
\usepackage{colordvi}
\usepackage[absolute]{textpos}
\textblockorigin{-6mm}{0mm}

\usepackage[
    colorlinks = true,
    urlcolor = blue,
    menucolor = blue,
    linkcolor = blue,
    citecolor = blue
]{hyperref}
\usepackage{booktabs}
\usepackage{dsfont}
\usepackage{wallpaper}
\usepackage{mathtools,cancel}
\usepackage{soul}
\usepackage{float}
\usepackage{setspace,titlesec}
\usepackage{amsmath,amssymb,amscd,bm,amsfonts}

\usepackage{amsthm,bm}
\usepackage{tabularx, graphics, longtable,makecell}
\usepackage[toc,page]{appendix}
\usepackage{verbatim} 
\usepackage{authblk}
\definecolor{green}{rgb}{0,0.6,0} 
\definecolor{red}{rgb}{1,0,0} 
\definecolor{blue}{rgb}{0,0,1}
\definecolor{LINK_COLOR}{rgb}{0,0,0.7}
\definecolor{CITE_COLOR}{rgb}{0,0.5,0}
\definecolor{light-gray}{gray}{0.95}
\usepackage{breakcites}
\usepackage{tocloft}
\usepackage{abstract}

\usepackage{subfig}

\titlespacing*{\section}{0pt}{0.5\baselineskip}{0.1\baselineskip}
\titlespacing*{\subsection}{0pt}{0.5\baselineskip}{0.1\baselineskip}
\titlespacing*{\subsubsection}{0pt}{0.5\baselineskip}{0.1\baselineskip}

\usepackage[font=small,skip=3pt]{caption}

\makeatletter
\def\@seccntformat#1{\@ifundefined{#1@cntformat}%
   {\csname the#1\endcsname\quad}  
   {\csname #1@cntformat\endcsname}
}
\let\oldappendix\appendix 
\renewcommand\appendix{%
    \oldappendix
    \newcommand{\section@cntformat}{\appendixname~\thesection\quad}
}
\makeatother

\title{\textbf{Cycles Protocol: A Peer-to-Peer Electronic Clearing System}}
\author{Ethan Buchman, Paolo Dini, Shoaib Ahmed, Andrew Miller, Tomaž Fleischman}
\date{
\vspace{-0.5cm}
November 2024 - v0.5 \footnote{Learn more at \href{https://cycles.money}{cycles.money}}
\vspace{-0.5cm}
}

\begin{document}
\maketitle

\begin{abstract}

For centuries, financial institutions have responded to liquidity challenges by forming closed, centralized clearing clubs with strict rules and membership that allow them to collaborate on using the least money to discharge the most debt. As closed clubs, much of the general public has been excluded from participation. But the vast majority of private sector actors consists of micro or small firms that are vulnerable to late payments and generally ineligible for bank loans. This low liquidity environment often results in gridlock and leads to insolvency, and it disproportionately impacts small enterprises and communities. 

On the other hand, blockchain communities have developed open, decentralized settlement systems, along with a proliferation of store of value assets and new lending protocols, allowing anyone to permissionlessly transact and access credit. However, these protocols remain used primarily for speculative purposes, and so far have fallen short of the large-scale positive impact on the real economy prophesied by their promoters.

We address these challenges by introducing Cycles, an open, decentralized clearing, settlement, and issuance protocol. Cycles is designed to enable firms to overcome payment inefficiencies, to reduce their working capital costs, and to leverage diverse assets and liquidity sources, including cryptocurrencies, stablecoins, and lending protocols, in service of clearing more debt with less money. Cycles solves real world liquidity challenges through a privacy-preserving multilateral settlement platform based on a graph optimization algorithm. The design is based on a core insight: liquidity resides within cycles in the payment network's structure and can be accessed via settlement flows optimized to reduce debt. 

\end{abstract}

\clearpage

\tableofcontents

\clearpage
\section{Introduction}
We live in a world of increasing financial inequality among firms, aggravated by growing requirements for collateral when accessing formal financing sources \cite{DeMarcoetal2021}. The vast majority of corporate actors consists of micro and small firms with zero or modest collateral, which reduces their eligibility for bank loans \cite{FinneganKapoor2023,ECB2023,BiaisGollier1997}. This aggravates the late payment problem \cite{Lefebvre2023,Berlocoetal2021,Walker2017,PetersenRajan1994,FabbriKlapper2016}, leading to gridlock \cite{Leinonen2005}, and often turning liquidity challenges into insolvency \cite{Boissay2006}.  Firms are forced to seek informal liquidity sources -- or in the worst case loan sharks -- to access the working capital they need for their normal operations and growth. The Great Financial Crisis (GFC), the recent Covid public health crisis, and on-going environmental degradation are making their situation even harder. The rapid development of alternative monetary and financial systems such as complementary currencies and cryptocurrencies over the past 15 years can be seen as a direct response to this situation.

It has always been known that sharing financial information can produce better outcomes. Clearing systems are a primary example. For centuries, banks and payment providers have improved their profitability and stability by forming closed clearing clubs with strict rules to make collaboration possible \cite{BoyerX,Bertschingeretal2019}. Clearing allows them to extinguish large amounts of debt using minimal amounts of money or liquidity in a coordinated and certain manner. Clearing institutions came to serve as a linchpin of risk management in financial systems, even though they, somewhat paradoxically, introduce new central counterparties to whom significant risk is transferred. However, most of the public is excluded from accessing these clearing systems. The general public and small firms cannot collaborate on clearing since there are no rules to protect them from the harm of exposing their financial data to their competitors and partners, and membership in clearing house clubs generally involves high-overhead financial contracts that put them out of reach. 

Recent advances in privacy-preserving technology, distributed consensus, and graph algorithms allow us to overcome these challenges. With privacy technology, debts can be securely and inexpensively collected from a large number of participants. With distributed consensus, debts can be cleared by the fault-tolerant execution of atomic multilateral operations, allowing for the simultaneous discharge of a large number of debts. And, with graph algorithms, debts can be cleared in a risk-reducing manner without transferring risk to central counterparties. In other words, we can now develop Cycles, a payment system that optimizes to clear the most debt by performing a large number of settlements in a single operation, benefiting a wide number of participants, without the introduction of intermediaries or financial complexity. This opens further use cases for lending and issuance protocols, and allows firms and communities to greatly improve their liquidity position and reduce risk.

Here, we describe a common language and framework for the design of payment systems, and we sketch an initial implementation of the Cycles Protocol. The language exposes the structure of the payment system as a network of obligations among agents and liquidity sources. The framework is based on a graph optimization algorithm which allows a large number of participants (debtors, creditors, and liquidity providers) to benefit from clearing the most debt with the least money. The protocol enables users to privately pool their obligations and their preferences over the use of available liquidity (cryptocurrencies, stablecoins, lending and issuance protocols, etc), and to execute optimized atomic multilateral settlements across users.

The design is motivated by a core insight: liquidity resides within the network structure of debts and can be accessed via cyclic settlement flows optimized to reduce debt. By surfacing the graph structure of the network in a privacy-preserving manner, operating on it atomically with multilateral settlement flows, and integrating diverse sources of liquidity, major benefits can accrue which are not otherwise accessible to individual companies, to trade networks, and to whole economies. 

With Cycles, firms can connect their internal accounting system to a global network that optimizes the clearing of credits and debts using the available sources of liquidity. Firms select which of their debts they want to submit to the clearing system, and what kinds of assets and credit sources (amounts, terms, etc.) they want to use to pay them. Our initial emphasis is on the trade-credit economy, and its network of accounts payable. These are typically 30, 60, or 90-day 
credits extended by suppliers to their customers (i.e.\ invoices), that only bear interest once they are overdue. They are a major source of the cash flows (and stresses) that dominate the operations of a business. In some sense they are the financial foundation of a commercial economy. Our design presents a practical solution that optimizes the clearing of this (and other) debt via simple legal patterns, improving cash flow, and reducing dependence on expensive sources of credit like factoring and bank loans. While trade credit is presented as an initial focus, the design of Cycles is more widely applicable to debts and payments in general (rents, wages, interest payments, etc.). It offers a new way to implement robust payment systems and a new foundation for reasoning about finance. 

In the first half of this paper, we develop the language of payments and the core design problem of graph optimization over a network of obligations and liquidity sources to achieve atomic multilateral settlement. In the second half, we sketch an initial protocol that implements the design and discuss some 
considerations and extensions. 

\subsection{Payment Systems}
\label{payment-systems}
We define a payment system to consist of a set of obligations (the debts to pay), and at least one liquidity source (typically, some asset) that can be used to discharge them. The obligations, together with offers to use and accept different sources of liquidity, form a network graph. Many of today’s payment services (banks, fintechs, blockchains) focus primarily on the transfer and exchange of assets, with limited support for obligations, and with a limited view of the network graph. But these existing systems become much more powerful when they are integrated into an obligation graph of the kind we propose. As we will see, first class representation of obligations and the ability to perform atomic multilateral operations unlocks powerful new capabilities for the collaborative discharge of debt.

Our design is motivated by a critique of existing payment systems, summarized briefly here. Our aim with this critique is to offer a direction forward for enhancing and complementing existing systems, rather than replacing them.  





First, the modern banking system was not derived from anything approaching a coherent theory of finance and economics, and has become quite fragile in its relationship to debt \cite{FragileByDesign}. By contrast, we seek to ground our proposed payment system on \emph{first principles}, by asking the question: ``How do we design a payment system to reduce the most debt with the least money for everyone?" 

Second, much of the risk management in modern systems is focused around central banks as lenders and dealers of last resort \cite{mehrling2011new}. While useful for backstopping certain kinds of liquidity crises, this structure has led central banks to be captured by systemic risk \cite{ozgode2021} and to compound moral hazard \cite{LeeLeeColdiron2020,chancellor2022price}. Instead, our design focuses on risk-reduction mechanisms accessible to the general public by enabling them to use a wider variety of assets and clearing protocols to make payments.

Third, much payments innovation is focused on the bilateral transfer and exchange of assets by individuals. However, the payment system has a \emph{network structure} arising out of the web of obligations formed in the course of trade and finance. Our proposed payment system allows this multilateral network structure to be surfaced and optimized over with the tools of graph theory. We focus on the network structure of the liabilities, rather than the aggregate structure of the assets.

Fourth, modern payment systems often invoke intermediaries, counterparty substitution, and contract novation, which are associated with a higher regulatory burden and transmutation of risk. By contrast, we can reduce the need for intermediaries and financial complexity by focusing instead on the existing network of liabilities (especially trade credit obligations) and leveraging the more permissible legal structure of set-off notices under international private obligation law \cite{Unidroit2016}. This approach honours the network of relationships in the obligation graph and allows the focus to remain on network-level risk reduction. It allows debts to be reduced by formal set-off transactions that reduce debts for multiple parties at once.

Fifth, liquidity in modern banking and blockchain systems is organized around a system of market-making dealer intermediaries focused on their own enrichment through \textit{liquidity provisioning} \cite{AmatoFantacci2012a}. In contrast, our proposed payment system is designed without such intermediaries, and is focused on \textit{liquidity saving} via set-off. Liquidity provisioning is a short volatility position associated with systemic risk.\footnote{Also known as `picking up pennies in front of the steamroller' \cite{LeeLeeColdiron2020}. In essence, betting on stable prices.} Liquidity saving is a way to reduce that systemic risk  \cite{FleischmanDini2021}.

Finally, payment systems must ultimately reckon with the problem of \textit{issuance}, which arises when there is not enough liquidity in the system. There is much to lament about modern issuance through commercial and central banks \cite{bagehot,minsky2008,TOMAC}. Our proposed design leverages the obligation network -- credit and debt relationships within the non-financial sector -- as endogenous network liquidity to enable new forms of distributed issuance that improve the system's overall liquidity. We thus present a platform for new kinds of credit and issuance protocols that are more ``network-aware''.

One way to understand our design is that it introduces a clear separation between two key functions of money: the Unit of Account and the Medium of Exchange. In today's monetary systems, these are often conflated in a single ``currency", though historically they were more likely to be distinct concepts~\cite{spufford}. For our purposes, the function of money is to make payments -- to denominate and discharge debts. The Unit of Account function, then, is for denominating debts, while the Medium of Exchange function is for discharging them, here and now.\footnote{Store of Value is often considered a third function of money. If the Medium of Exchange function is for discharging debt \textit{here and now}, the Store of Value function is for discharging debt \textit{elsewhere or later}.} The graph of obligations at the heart of our design is a pure expression of the Unit of Account function, while our objective of clearing that debt using the least amount of ``money" is an attempt to leverage any conceivable source (including the debts themselves!) for the Medium of Exchange function -- realizing that debts can be cleared using any asset, or without even using any money at all. This separation of functions within the design of the payment system allows for much greater efficiency and more diversified expression of value. 

Our design emerges from a synthesis of traditions which are at their core efforts to implement robust and sustainable payment systems.\footnote{See prior work on obligation-clearing for liquidity-saving \cite{SimicMilanovic1992,EisenbergNoe2001,Fleischmanetal2020} with mutual credit  \cite{Studer1998,Greco2009,LietaerDunne2013,SartoriDini2016,Litteraetal2017,DiniKioupkiolis2019} and the simultaneous use of multiple liquidity sources to discharge debt.}
This synthesis is driven by the observation that a great deal of financial value in support of the real economy is not visible to selfish rational agents reasoning about their own assets, and can only be accessed through collaborative processes that allow us to operate over the network of liabilities, which we share an interest in discharging. 
We seek to harness these collaborative processes to reduce the constant liquidity pressure currently bearing down on the global economy.

Thus, without rejecting competition -- or capital markets more generally -- we uncover a new source of resilience and sustainability for networks of real economy actors, especially small and medium-sized enterprises (SMEs), and new ways to leverage internal and external liquidity for mutual benefit. In so doing, we offer a new path to empower communities to manage their own payment systems and to issue their own money in a sustainable fashion. In turn, new possibilities emerge for existing pools of capital to support a more sustainable finance.

\section{Design}
\label{design}
Our goal is a universal language for representing arbitrary financial relationships as multilateral graph settlement operations on a network of balance sheet T-accounts.  We introduce terminology and a graphical schema to describe a wide variety of payment, currency, and credit systems as a network of debt and credit relationships between balance sheets. This language motivates a graph optimization problem whose solution can be executed as a single multilateral operation, a settlement flow that discharges many debts with minimal liquidity. The language is based on a system of intents that allows participants to express their indebtedness and their preferences over the use of different liquidity sources.

As we've noted, much of modern finance today revolves around a bilateral \textit{transactional} view, focused on individual settlements, in contrast to our multilateral \textit{network} view, focused on settlement flows across a network. The transactional view misses countless opportunities for liquidity saving made possible by the network view. Further, the transactional view tends to invoke the additional legal constructions of counterparty substitution and/or contract novation (securitization, factoring, clearing houses, etc.), which implies replacement of an obligation or counterparty. Changing contracts or counterparties is expensive, slows down business, and mutates risk. We propose a design that limits the need for contract novation and counterparty substitution by instead making maximal use of the \textit{existing graph structure}\footnote{Respect the Graph.} of obligations and credit relationships. This has the added legal benefit of moving from the complex world of financial regulation to the simpler world of private obligation law, which is more accessible and better able to adapt to different contexts. This framing allows us to also pursue novel use cases for lending and issuance protocols, empowering communities to make better use of these monetary powers in a more distributed and sustainable fashion. 

In the following sections we lay out the foundations for the network view. We start with a graphical intent system and a complementary balance sheet view of the four possible ways to settle. The network of intents is operated on by graph flow algorithms, resulting in balance sheet liquidity savings for a combined network settlement system.

\subsection{System of Intents}

Our system is based on two key intent types which we call Obligations and Acceptances, an overlay type called Tenders, and an output type called Settlement Records. Obligations and Acceptances are commitments to past and future debts, respectively, while Tenders initiate settlement and Settlement Records denote how much of a tendered obligation or acceptance is to be discharged. Generally speaking, any payment transaction can be decomposed into these components. We assume intents are programmable, allowing a wide variety of payment, currency, and credit protocols to be defined. 

\textbf{Obligations.} An obligation is the core of any payment -- it's the debt the payment settles. If Alice owes Bob \$30, that's an obligation. More generally an obligation is a tuple consisting of a debtor, creditor, and amount. Obligations originate from the debtor as a declaration of their liability to the creditor. An obligation is an intent to pay. It must be \textit{ascertained} (signed or accepted) by its debtor, but it could technically require no ascertainment from the creditor. If Alice declares she owes Bob, who is Bob to stop her? 

Many common obligations also include a due date: invoices, rent, interest, wages, etc. Both debtors and creditors of obligations with due dates have a general incentive to see that the obligations are paid, or \textit{discharged}. Obligations can be fully or partially discharged. Full discharge is usually referred to as settlement. Our design allows the problem of full discharge to be posed as a graph optimization, and it also allows agents to take advantage of the benefits of partial settlement.

Creditors might also wish to accumulate obligations from their best debtors, obligations they don't necessarily want paid yet, especially if they can call them on-demand. For instance, cash balances at the bank, which are obligations from the bank that you can get the bank to honour at any time. More generally, possession of \textit{any} digital asset can be understood as an obligation from the asset's issuer (i.e.\ a bank, a blockchain, etc.) to the asset holder.

\textbf{Acceptances.} In its most basic form, an acceptance defines how someone wants to be paid, and thus determines how an obligation can be discharged. Acceptances originate from creditors. If Alice owes Bob, and Bob says, ``I'll only accept bank money as payment", that is an acceptance from Bob to bank money (we might also say an acceptance ``for" bank money).

\begin{figure}[H]
\centering
{\includegraphics[width=11cm]{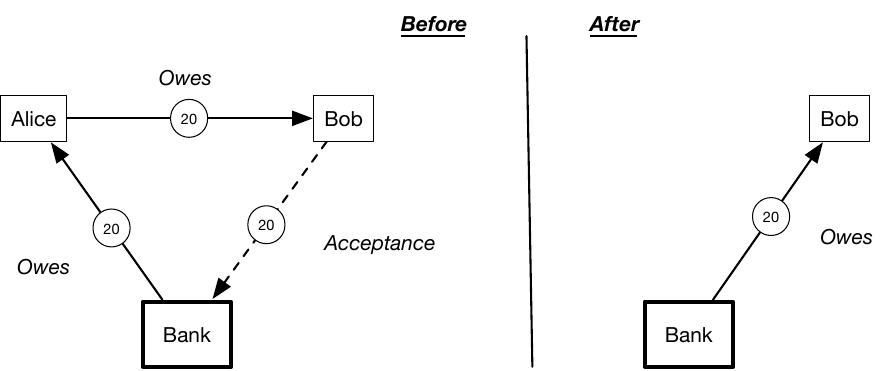}}
\caption{\small\textbf{Language of Obligations and Acceptances}. Solid arrows are obligations, dashed arrows are acceptances. The bank owes Alice (cash deposits), who owes Bob (e.g. an invoice), who is willing to be owed by the bank (an acceptance of bank deposits). After executing a payment, Bob's acceptance is replaced by an obligation in the other direction, and the other obligations are discharged. By accepting to the bank, Bob is ``lending" his money to the bank. Instead of being owed by Alice, Bob would rather be owed by the bank (checking account). In Cycles, the Bank could be any digitally programmable liquidity source -- for instance, BTC, or a smart contract.}
\label{fig:obligation_acceptance}
\end{figure}

But the settlement of an acceptance immediately creates an obligation in the opposing direction, as shown in Fig~\ref{fig:obligation_acceptance}, which defines the basic language of obligations and acceptances. When Bob gets paid, his \textit{acceptance to} the bank becomes an \textit{obligation from} it. Bob will generally have an infinite size acceptance to the bank (willing to be owed an infinite amount by the bank), especially if bank deposits are legal tender. But he may have a smaller acceptance to other currency issuers. Of the obligations owed to him, maybe he'll only accept up to \$1000 paid in BTC, while the rest must be paid in bank deposits.

We can see right away from Fig.~\ref{fig:obligation_acceptance} that \textit{every payment is actually a cycle} when you include the liquidity source. It's common to think about accepting money to the bank as creating an obligation from the bank back to you. But ownership of any digital asset can be represented fruitfully as an obligation from the asset's ``issuer''. Hence even self-custodied BTC assets can be represented as obligations from the Bitcoin blockchain. If you own BTC, the blockchain ``owes'' you the right to move it. Thus replace Bank with Bitcoin in Fig~\ref{fig:obligation_acceptance}. Representing asset balances as obligations from liquidity sources allows for a much richer expression of the payment graph.

Acceptances then are commitments to future obligations -- commitments from banks and the Bitcoin blockchain to owe others, and even to owe them on-demand (you can use your cash or BTC at any moment). We tend to call agents in the economy with a large number of such on-demand acceptances directed to them \textit{liquidity sources}. They're the agents everyone most wants to be owed by, and they're the agents willing to owe everyone. We call an acceptance like this, that spawns an on-demand obligation in the other direction, a Deposit Acceptance.

But suppose Alice doesn't have cash and actually needs to borrow money from the bank to pay Bob. Bob still has an acceptance \textit{to the bank}, indicating his willingness to effectively ``lend" to the bank. But if the bank is willing to lend to Alice, this too can be represented as an acceptance, only \textit{from the bank} to Alice. The main difference is that Alice's loan from the bank will typically have a repayment date, while Bob's loan to the bank is available on-demand (a deposit).\footnote{Of course both might involve an interest rate, which generates additional interest payment obligations. Interest is a higher order concept that can be layered on top via the system's programmability.} We call an acceptance that creates an obligation with a repayment date a Repayment Acceptance. Deposit acceptances can be made to any kind of liquidity source and repayment acceptances can be made from any kind of a lending facility.

So obligations are commitments to debts in the past while acceptances are commitments to debts in the future. How are settlements actually initiated? Enter tenders.

\textbf{Tenders.}
In its most basic form, a tender defines how someone wants to pay an obligation. Alice might owe Bob \$30, and she might have both bank deposits and a stablecoin like USDC. If Alice says, ``Will you accept this 30 USDC to settle my debt?", that is a tender. More generally, a tender is an intent to use up to some amount of a particular source of liquidity to discharge obligations you owe. 

If Alice has gold, dollars, and BTC, she might tender any of them, or all three, to pay her debts. For a non-stable liquidity source like BTC to settle a debt in dollars, the tender must specify an acceptable price for BTC in dollars. This price could come from an oracle, or it could be specified by Alice. Note the BTC are not actually exchanged for dollars, but they are used to pay a dollar denominated debt.

When you tender something you already have possession of, like a bank deposit or a stablecoin, the tender draws on a pre-existing obligation. Your bank deposit is the bank's liability, and your BTC is the Bitcoin blockchain's liability. More generally, we call what you have possession of a positive liquid balance, and we call whatever owes it to you a liquidity source. Any positive digital balance can be understood as an obligation from a liquidity source.

But you can also tender assets you don't already have (for which there is no pre-existing obligation), if someone is willing to lend them to you or if you are authorized to issue them -- in other words, if there exists a repayment acceptance to you that you can draw on. A bank extending a \$1000 credit line to Alice can thus be understood as a \$1000 acceptance from the bank to Alice, which she can tender from to pay her obligations. 

Once a set of obligations and acceptances are tendered and ascertained, they can be settled by applying a settlement record.

\textbf{Settlement Records.} Settlement records are the output of the graph flow algorithm applied to a network of tendered obligations and acceptances. They quantify how much of each obligation or acceptance is to be discharged. Settlement records must be applied atomically in valid batches called Settlement Flows. A settlement flow is a balanced cyclic set of settlement records -- every valid payment is a cycle. In Fig.~\ref{fig:obligation_acceptance}, the transition from \textit{before} to \textit{after} is triggered by the application of the settlement records in the settlement flow. For each edge (obligation or acceptance), a settlement flow contains two settlement records, one for each node (for the same edge, e.g. an invoice). And for each node, a settlement flow contains at least two records, balancing flows in and out (e.g. for offsetting two different invoices).

From a network perspective, the settlement records in a cycle must execute atomically, since they are mutually dependent. Either they all apply as a cycle, or none of them do. In the case of a simple payment of an invoice by bank transfer from Alice to Bob, the bank's internal settlement system is sufficient to discharge Alice's obligation to Bob. But to represent more complex networks and settlement flows requires a platform for atomic execution of settlement records -- namely, a blockchain.

In the case of an on-chain liquidity source like USDC or BTC, settlement records indicate balance changes up or down for a user. But in the case of off-chain assets like invoices (even if represented on-chain), a settlement record must generate a corresponding \textit{set-off notice}, which we define as a formal and legally binding communication about the result of a settlement process that can be applied to a balance sheet. While settlement records can be executed on-chain, set-off notices must be executed by independent actors off-chain (i.e. by marking down receivables and payables in local accounting software). 

By describing payments in terms of obligations and acceptances with overlaid tenders operated on by balanced settlement records, we can optimize to find a settlement flow that discharges the most debt with the least (and most preferred) liquidity. The presence of cycles and chains of obligations, both across agents and liquidity sources, allows the network to discharge more debt than would be possible otherwise, thanks to the power of set-off and the different ways to settle.

\subsection{Four Ways to Settle}
\label{4ways}

\begin{figure}[H]
\centering
{\includegraphics[width=16cm]{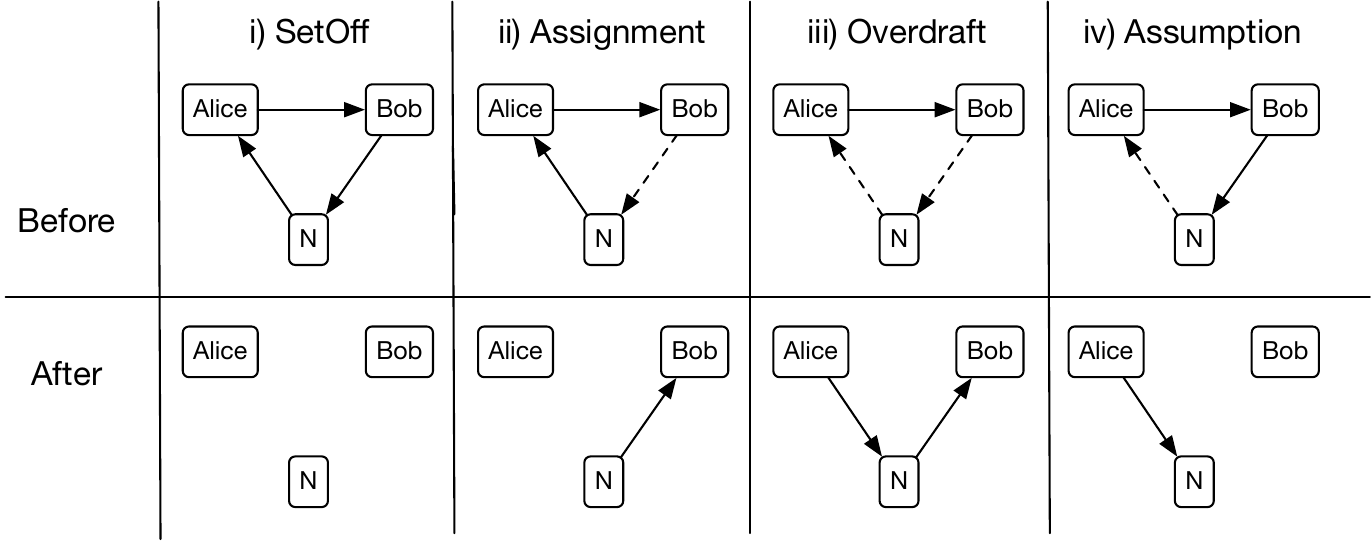}}
\caption{\small\textbf{Four Ways to Settle}. There are four ways to settle the debt from Alice to Bob based on the four combinations of assets and liabilities from Alice's and Bob's balance sheets. Either Alice has assets (i, ii) or not (iii, iv) and either Bob accepts an asset (ii, iii) or reduces one of his liabilities (i, iv). Graphically, these correspond to the four possible configurations of obligations and acceptances going into Alice and coming out of Bob, completing a cyclic flow. Every settlement is a cyclic flow. We use N to represent the rest of the network through which the settlement flows. In these examples, N is most simply understood as a liquidity source (like a bank), but it could be any arbitrary network. Note that assignment and assumption are rotations of one another -- what is assignment for Alice is assumption for N, and assumption for Alice is assignment for Bob.}
\label{fig:4ways}
\vspace{-0.4cm}
\end{figure}

We now combine the graphical language of intents with the common accounting language of balance sheets. 
In Fig~\ref{fig:obligation_acceptance} we saw Alice settle a debt by transferring an asset to Bob. But from a basic balance sheet perspective, there are four ways to settle an obligation \cite{Clavero2022}, which we refer to as (i) set-off, (ii) assignment, (iii) overdraft, and (iv) assumption.\footnote{We are building on Clavero's formulation \cite{Clavero2022}. The four ways to settle are fundamental to bookkeeping and are given by the 2x2 matrix of whether debtor and creditor altered their assets or liabilities. While Clavero focused on describing accounts from the perspective of the payment system, we are more interested in a network perspective, so our terminology differs slightly. What is called \textit{issuance} in \cite{Clavero2022}, we call \textit{overdraft}, which in our formulation includes issuance as a special case (when new monetary units are created).What is called \textit{novation} in \cite{Clavero2022}, we call \textit{assumption}, to distinguish it from the legal definition of contract novation and to emphasize its symmetry with assignment.} As we'll see, they correspond to (i) balance sheet reduction, (ii) asset transfer, (iii) balance sheet expansion, and (iv) liability transfer. To summarize, if I owe you and you owe me, we can do (i) \textit{set-off} and reduce our balance sheets. If I have some assets, I can do (ii) \textit{assignment} and transfer the assets to you (Fig~\ref{fig:obligation_acceptance}). If I don't have assets, I can do (iii) \textit{overdraft} to borrow assets from someone else and transfer them to you, expanding some external balance sheet. Or if you have a liability, I can do (iv) \textit{assumption} to assume that liability from you. The four ways are shown graphically in Fig.\ \ref{fig:4ways}. 

These four ways to settle can be used to describe a wide variety of payment and financial flows, as we'll describe below. They can also be combined in arbitrary ways to enable more efficient network settlement. In what follows, we use balance sheets and graph representations to review the different ways to settle. This framing will allow us to more completely define the graph optimization problem at the heart of our payment system design. In each case, settlements are executed by applying settlement records to a network of obligations and acceptances. 

\begin{figure}[H]
\centering
{\includegraphics[width=11cm]{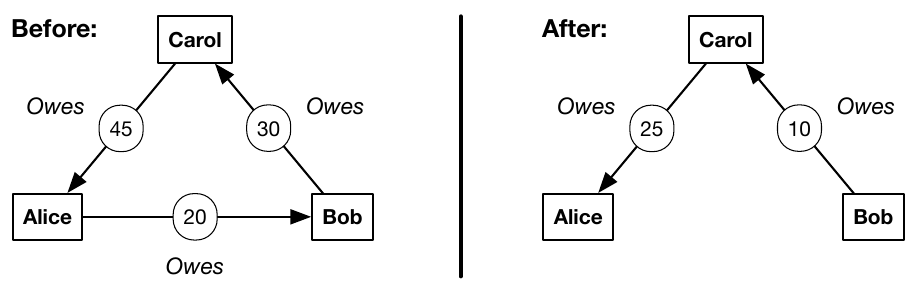}}
\caption{\small\textbf{Graph of 3-cycle set-off}. Alice owes Bob 20, who owes Carol 30, who owes Alice 45. With set-off, each obligation can be reduced by 20, fully discharging Alice's debt to Bob, and partially discharging the others. Total debt in the system drops from 95 to 35.}
\label{fig:setoff}
\end{figure}

\begin{figure}[H]
\centering
{\includegraphics[width=14cm]{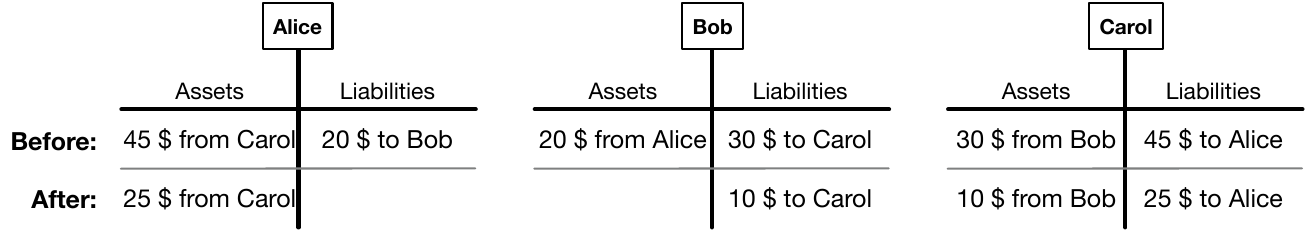}}
\caption{\small\textbf{Balance sheets for 3-cycle set-off}. The same as Fig.\ \ref{fig:setoff}, but shown as balance sheets, where obligations are liabilities for the debtor and assets for the creditor.}
\label{fig:bs_setoff}
\end{figure}

\textbf{Set-off.} 
Set-off is the discharge of obligations without money. This is done by balancing obligations across balance sheets so they offset each other. If Alice owes Bob and Bob owes Alice, they can do set-off. Set-off is more interesting when there are cycles of size greater than two -- if Alice owes Bob and Bob owes Carol and Carol owes Alice, they can all set off the lowest amount. We show this graphically in Fig. \ref{fig:setoff}, and using balance sheets in Fig. \ref{fig:bs_setoff}.

Set-off reduces the size of each party's balance sheet by the amount of the settlement record.

\begin{figure}[H]
\centering
{\includegraphics[width=14cm]{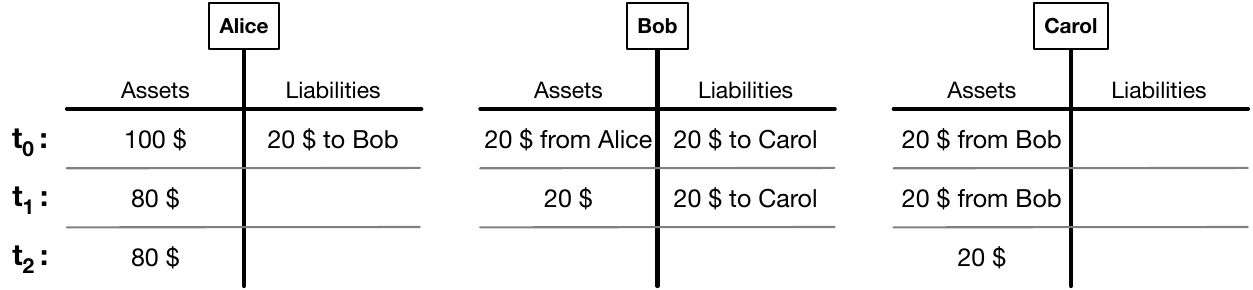}}
\caption{\small\textbf{Assignment chain}. Alice has \$100, and owes \$20 to Bob, who owes \$20 to Carol. Alice uses \$20 to pay Bob, who then pays Carol. For simplicity, all debts are the same size.}
\label{fig:bs_chain}
\end{figure}

\textbf{Assignment.}
Assignment is the discharge of an obligation by the transfer of assets. Unlike set-off, which is multilateral, assignment can be initiated unilaterally by the debtor of an obligation. Multiple consecutive assignments allow the same money to be used to discharge a chain of multiple obligations. Fig.\ \ref{fig:bs_chain} shows a chain of assignments involving three parties transferring \$20 to clear \$40 of debt. Assignment only reduces the balance sheet of the debtor via an asset transfer that reduces their liability. For the creditor, assignment is asset substitution (i.e.\ replacing a receivable with cash).

\vspace{-0.4cm}

As we saw in Fig~\ref{fig:obligation_acceptance}, it is useful to think of a liquidity source as a node in the graph, which can be connected to debtor and creditor nodes via obligations and acceptances, respectively. The liquidity source, in the case of assignment, is the keeper of positive balances, its obligations to the firms. The liquidity source could be a bank, a blockchain, or a mutual credit system. When representing obligations from liquidity sources, we restrict them to only showing the amount tendered. Alice could have \$100 in the bank (an obligation of \$100 from the bank), but only tenders \$20.

If the liquidity source is part of an atomic settlement network, like a blockchain, then assignment and set-off can be combined atomically into a multilateral operation so that assignment (the asset transfer) only occurs between the first and last nodes in a chain of obligations, while all other obligations in the chain are set off (i.e.\ going direct from $t_0$ to $t_2$ in Fig.~\ref{fig:bs_chain}). This avoids multiple independent assignments down the chain, achieving the same result but in one step, and without creating a new relationship between the first and last parties in the chain. This is important because it allows for significant savings and optimizations over the graph without counterparty substitution or contract novation.

\begin{figure}[H]
\centering
{\includegraphics[width=12cm]{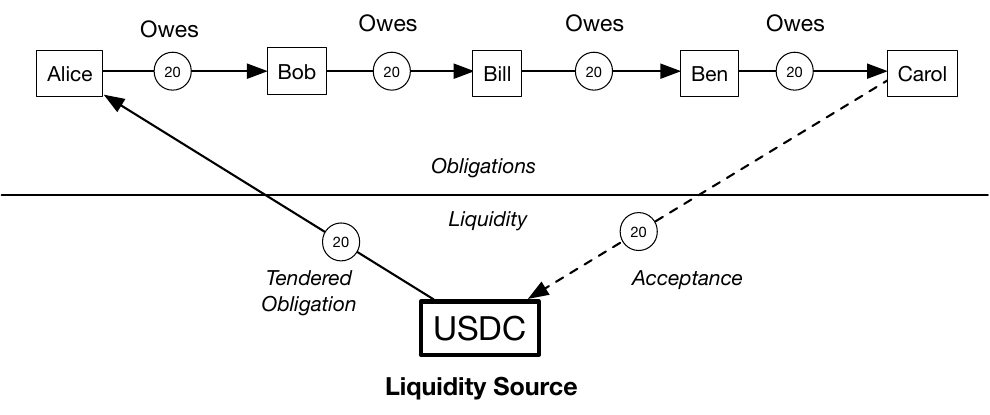}}
\caption{\small\textbf{Multilateral Settlement via Assignment}. Many debts can be cleared by combining a single assignment with many set-offs. Alice's money gets assigned to Carol, and all debts are set off, without any new relationship between Alice and Carol. The liquidity source could be any digitally programmable asset, like USDC, BTC, etc.}
\label{fig:liquidity_assignment}
\end{figure}

We depict the multilateral case in Fig.\ \ref{fig:liquidity_assignment}. Alice owes Bob, who owes Bill, who owes Ben, who owes Carol. Bob's, Bill's, and Ben's debts are cleared by the transfer of 20 units of liquidity directly from Alice to Carol, atomically and with set-off notices received by all of them. In a single operation all the debts are discharged and Alice's 20 is transferred directly to Carol, without Alice or Carol ever knowing about each other. Bob, Bill, and Ben never handle any money, and all their debt is cleared by set-off. All the obligations disappear, and Carol's acceptance to the liquidity source becomes an obligation from it. 


\vspace{-0.4cm}

In general, we will refer to tendering obligations from a liquidity source as assignment tenders or just assignment.

\textbf{Overdraft.}
\label{overdraft}
Overdraft is discharge of one obligation by the creation of a new one -- borrowing to pay. While assignment draws on an obligation (an existing asset), overdraft draws on an acceptance (a new debt). The dynamics of the new debt, its credit limit, interest rate, and repayment, can be defined by a lending protocol, or what we call an overdraft facility. For a given type of liquidity, there can be many different overdraft facilities, each defined by its own lending protocol or capital pool. In general, we will refer to tendering such repayment acceptances as overdraft tenders or just overdrafts.

\begin{figure}[H]
\centering
{\includegraphics[width=16cm]{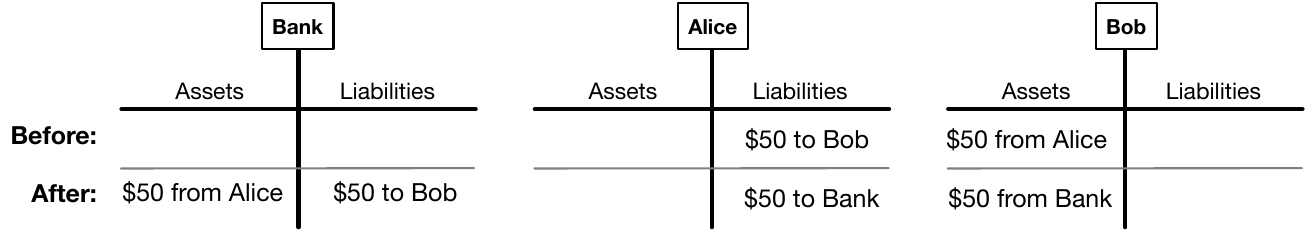}}
\caption{\small\textbf{Overdraft}. Alice draws on an overdraft facility (a line of credit with the Bank). Alice and Bob's balance sheets stay the same size, while the Bank's balance sheet expands.}
\label{fig:bs_overdraft}
\vspace{-0.4cm}
\end{figure}

Fig.\ \ref{fig:bs_overdraft} shows a common case. A Bank opens an overdraft facility for Alice. Alice owes Bob, and Bob is happy to accept bank money. Alice can draw on this overdraft facility, running up a new debt, which appears as an asset to the Bank and a new liability to Alice, and which can be used to pay Bob. In the end, Bob has bank money, and Alice owes the bank instead of Bob. Unlike set-off and assignment, overdraft alone does not reduce the size of Alice or Bob's balance sheets, but it does \textit{increase} the balance sheet of the Bank.

This bank balance sheet expansion is money creation, or \textit{issuance}, and reflects the primary way new money is created in the economy: banks extending credit by issuing new deposits. Note that the bank is not using money it already has. And note also that Alice does not first take the loan, and then use the asset she acquired in the loan to pay Bob via assignment -- taking on the new debt and paying the old debt are a single action. This is the meaning of an overdraft facility as used for payments (i.e.\ a debt used to reduce other debt), in contrast to a loan (i.e.\ a debt used to acquire an asset). And it's possible because Bob accepts the bank's liabilities as money. So the overdraft facility and the liquidity source are the same node -- the Bank.

\begin{figure}[H]
\centering
{\includegraphics[width=12cm]{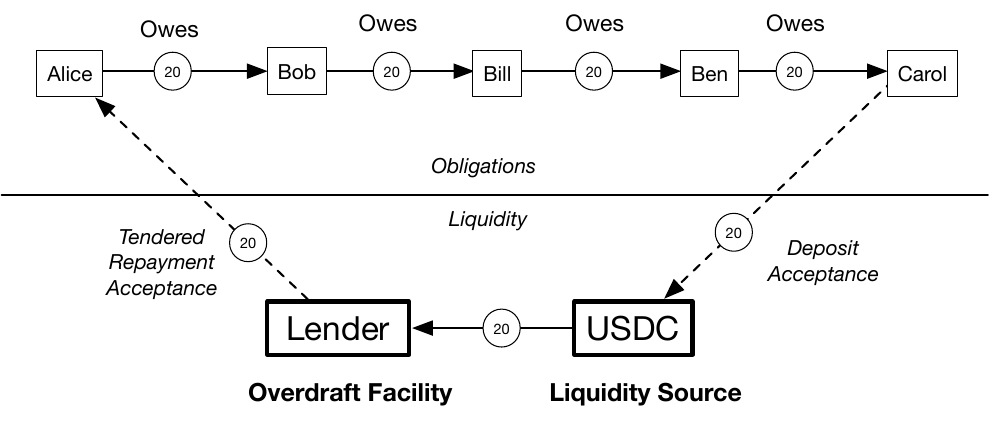}}
\caption{\small\textbf{Multilateral Settlement via Overdraft}. Alice doesn't have money, but a Lender does, and extends an acceptance to her. By Alice drawing on this credit line, Carol gets paid and all other debts are set off. This is the same as Fig.\ \ref{fig:liquidity_assignment}, except now Alice will owe the Lender.}
\label{fig:liquidity_overdraft}
\end{figure}

But what if Alice doesn't want to borrow from the Bank, and prefers a local Lender? The lender can't create money, and Bob doesn't want to hold money with them (they're not a bank), so Alice needs to take a loan, and then pay. But with multilateral settlement, any lender can provide a credit line through such an overdraft facility. When the overdraft facility and liquidity source are separate nodes like this, we refer to it as \textit{overdraft without issuance}, and we depict it graphically as in Fig.\ \ref{fig:liquidity_overdraft}. While the balance sheet of such a facility does not expand (they're lending existing assets), the combined balance sheet of the overdraft facility and the liquidity source does expand.\footnote{This is a subtle point that underlies why we use the term \textit{overdraft} to refer to what is called \textit{issuance} in \cite{Clavero2022}. We reserve the term \textit{issuance} for a network level distinction. In Fig.\ \ref{fig:4ways} we represented everything outside Alice and Bob as a single node N. This N ``provides'' the liquidity to settle Alice's debt to Bob, but can in itself be a complex network of nodes. The simplest case is that it is one node, a Bank. In that case, we'd say the overdraft facility and liquidity source are the same node, and thus it's overdraft \textit{with issuance}. But if the node N consisted of separate Lender and Bank nodes, where the Lender has money in the Bank, then the overdraft facility and liquidity source are separate nodes, and  it is overdraft \textit{without issuance}.}


\vspace{-0.4cm}

This distinction between overdraft with and without issuance enables us to think of issuance as a kind of credit, a negative balance expected to be paid back. Most money issuance today is done by expanding the balance sheets of commercial banks, using central banks as a backstop. But the network structure surfaced by our design makes new opportunities for issuance possible. Effectively, in our design, every agent can become a liquidity source, and every liquidity source an agent.

\textbf{Assumption.}
\label{novation}
Assumption is the discharge of an obligation by the transfer of a liability from one party to another. It is the inverse of assignment. Assignment transfers an asset from debtor to creditor (the debtor \textit{assigns} the asset), assumption transfers a liability from creditor to debtor (the debtor \textit{assumes} the liability). Like assignment, assumption reduces the balance sheet of one party while leaving the size of others unchanged.

As assignment's inverse, assumption is routine for payment providers. This follows from the symmetry in Fig.\ \ref{fig:4ways}. When Alice's obligation to Bob is discharged by assignment of her asset to Bob, the Bank's obligation to Alice is discharged by assumption of a new liability to Bob. Assumption is less familiar to regular businesses, though it can appear in the form of legal contract novation, where three parties agree to discharge one liability and replace it with another. A more common transaction, especially in trade credit markets, is factoring, where the creditor of an obligation is changed -- the original creditor sells the debt to someone else, usually at a discount. This is technically not assumption, but rather a form of overdraft. Exploring a few scenarios here will help illustrate some important differences between means of settlement that achieve similar results.

\begin{figure}[H]
\centering
{\includegraphics[width=17cm]{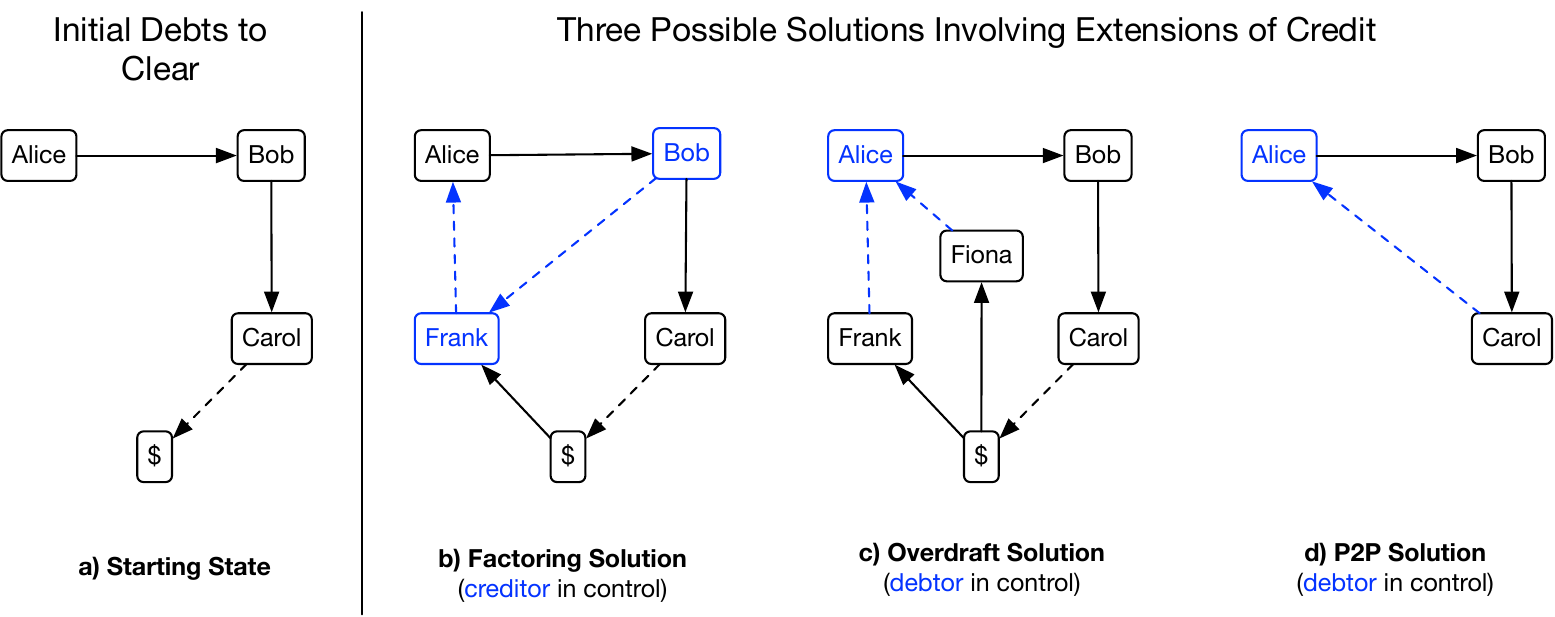}}
\caption{\small\textbf{Factoring vs Overdraft vs P2P Lending.} A comparison of three different solutions to getting Carol paid when (a) Alice owes Bob owes Carol. Blue indicates who is in control. Under (b) Factoring, Alice is forced to owe Frank, who paid Bob. Under (c) Overdraft, Alice can choose to borrow from (and owe) Fiona instead, who has no relation to Bob. Under a (d) P2P loan, Alice can choose to borrow from (and owe) Carol, who doesn't even need to advance assets. Factoring (b) is counterparty substitution and introduces Frank between Alice and Bob (Alice will owe Frank, who paid Bob). Note that Alice has no control over who Frank is. Frank is interested because he can purchase the receivable at a discount (not shown). The existence of discounting necessitates an extra acceptance from Bob to Frank and requires the settlement take place in two separate steps. In contrast, (c) Overdraft facilities and (d) P2P loans avoid counterparty substitution by using repayment acceptances and set-offs in a single operation. After settlement, discharged repayment acceptances become obligations from Alice (to Frank in (b), Fiona in (c) and Carol in (d)). In the case of (d), no assets are required, and the cycle clears, leaving only Alice owing Carol.}
\label{fig:novation_p2p}
\end{figure}

Suppose Alice owes Bob, who owes Carol, who accepts $\$$, as shown in Fig.~\ref{fig:novation_p2p}a. In invoice factoring (Fig.~\ref{fig:novation_p2p}b), Bob would factor (``sell'') his receivable from Alice to Frank for cash to pay Carol. In this case a liquidity provider, Frank, is buying the receivable from Bob, seeing it as an asset he can purchase at a discount and later collect on from Alice. This changes counterparties, since instead of owing Bob, Alice now owes Frank, whom she doesn't even know. Alice would much prefer to make the decision for herself, rather than be forced by Bob. She could consider borrowing directly from Frank, but maybe she'd prefer to borrow from Fiona (Fig.~\ref{fig:novation_p2p}c). This distinction is important.

\vspace{-0.4cm}

By framing the availability of liquidity in terms of the structure of the debts, rather than factoring or securitizing the assets, we can unlock new sources of liquidity within the network and empower debtors with the credit most appropriate for them. A powerful example of this is a `p2p loan', as shown in Fig.~\ref{fig:novation_p2p}d. Instead of borrowing assets from Frank or Fiona (outside the network), Alice borrows in the form of an acceptance from Carol, allowing the whole network to be cleared without any assets, leaving only a single debt from Alice to Carol to be paid in the future. Carol thus becomes a kind of liquidity source for Alice, hinting at a larger equivalence that emerges from thinking in terms of the graph: firms are liquidity sources and liquidity sources are firms.

Our goal here is to move from the common view of bilateral transactions to a new network view of multilateral settlement. While we can zoom in on a particular obligation and talk about set-off, assignment, overdraft, and assumption, we are ultimately concerned with larger networks and settlement flows. The point is, liquidity is in the graph! Starting with the obligation network, the addition of acceptances and liquidity sources, including p2p loans, extends the graph to make it more dense and cyclical. We can do clearing without counterparty substitution or contract novation because liquidity is hidden in the cycles. The key to our design then is an algorithm that finds cycles. 

\subsection{Graph Solving}
\label{mtcsalgo}
Any graph algorithm to find cyclic structures can generate valid settlement records and function as a \textit{solver}. The sets of settlement records generated by solvers must execute atomically. Our default solver is a min-cost max-flow algorithm called Multilateral Trade Credit Set-off (MTCS) \cite{Fleischmanetal2020,FleischmanDini2021}.

The existence of a cyclic structure implies that the amount of liquidity required to discharge all the debt is less than the total amount of debt. We call this amount the Net Internal Debt (NID). MTCS finds the minimum cost flow of this maximal amount of liquidity (the NID). Subtracting that flow from the full graph yields the cyclic structure. The mathematical details are described in \cite{FleischmanDini2021}. 

MTCS is designed for neutral settlement, but the optimal solution of the min-cost max-flow algorithm is not unique. There are in general many optimal paths satisfying the problem. While randomness could be used to pick between solutions, it would be better to enable other solutions to be expressed, for instance through governance, a preference system, or other parameters.

\begin{figure}[H]
\centering
{\includegraphics[width=11cm]{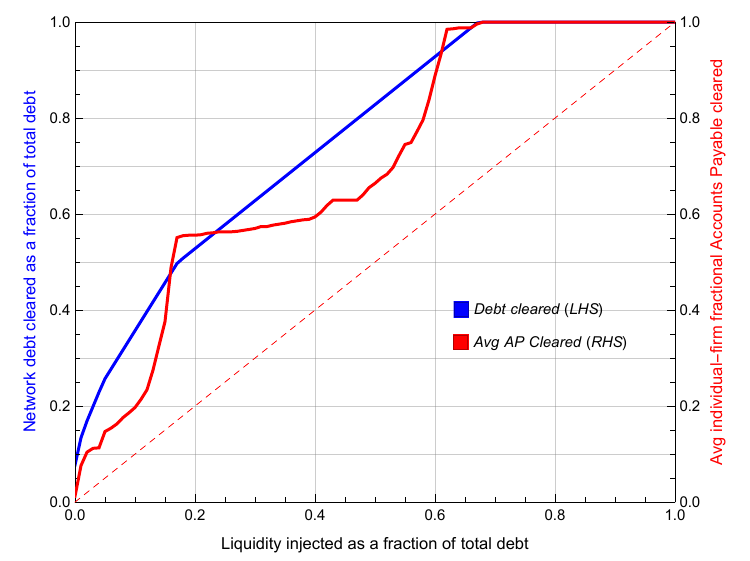}}
\caption{\small\textbf{Variation of debt set-off expressed as a fraction of total debt.} Percentage of debt cleared (blue) and average fraction of accounts payable (AP) cleared for each firm (red) plotted against the amount of liquidity injected as a fraction of the total debt. With no liquidity, nearly 10\% of the debt is in cycles and can still be cleared. The debt cleared (blue) rises steeply at first with slope greater than 1 (corresponding to chains of obligations), then grows with slope = 1 (corresponding to the clearing of isolated obligations), before levelling off at a fraction of injected liquidity that is significantly smaller than the total debt. The average accounts payable cleared (red) also grows with injected liquidity, with the first plateau corresponding to the clearing of large debts after the chains have been exhausted. The plot is based on anonymized Italian data: 1,280,000 invoices, 760,000 companies, December 2020.}
\label{fig:multiplier}
\end{figure}

Without a source of liquidity, only obligations in cycles can be cleared, and they can only be partially discharged (up to the smallest debt in the cycle). But even small amounts of liquidity can result in much greater amounts of debt being cleared, and with benefits for a larger number of participants. By adding at least the NID worth of liquidity, 100\% discharge of \emph{all} obligations is possible. The optimal solution found by MTCS implies that the amount of liquidity required is much smaller than the total debt, due to the simultaneous set-off of chains of obligations. Fig.\ \ref{fig:multiplier} shows an example of this `multiplier' effect.

\vspace{-0.4cm}

A production system may have diverse solvers, optimized for different outcomes, and with access to different views of the network and different liquidity sources. As we've seen, liquidity sources, lending protocols, and issuance protocols can be incorporated as nodes within the graph being optimized over by solvers, and can include the assets of debtors or external lenders, mutual credit, DeFi protocols, currency issuers, and even p2p loans. By exposing the graph to multilateral solving and thus enabling collaborative action by stakeholders, new forms of liquidity, and liquidity saving, become possible.

\subsection{Liquidity}

While a single liquidity source already provides significant benefits, we can introduce multiple sources of liquidity for the same network, compounding the opportunities for, and the overall volume of, setttlement flows, and greatly enhancing the network effect. Crucially, each currency ``circuit'' operates separately, so no currency exchange service is needed for settlement.

\begin{figure}[H]
\centering
\includegraphics[width=13cm]{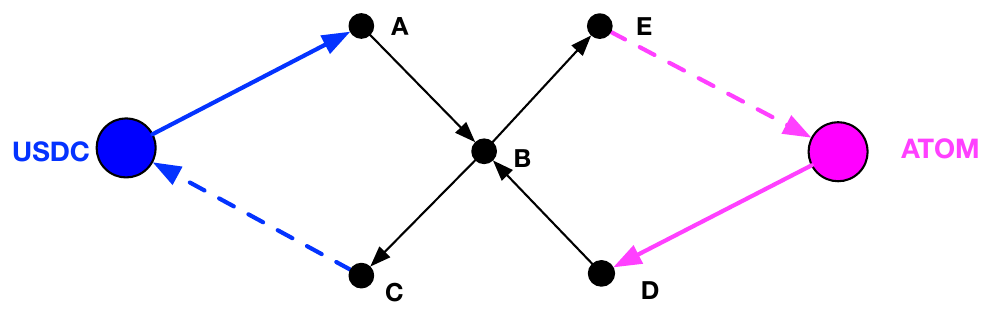}
\caption{\small\textbf{Two uncoupled liquidity sources}. B benefits from both without having to handle either.}
\label{fig:liquidity_uncouplediquidity}
\end{figure}

Consider the case of two uncoupled liquidity sources, shown in Fig.\ \ref{fig:liquidity_uncouplediquidity}. Assume for simplicity that all obligation amounts are equal and denominated in USD, and that we have access to a price oracle for all currencies in USD. We can see that Firm B is at the intersection of two separate cycles for payment in USDC and payment in ATOM. B will benefit from the set-off in each cycle separately, with no interaction between them, and without having to actually use either USDC or ATOM. In this case, B only engages with set-off notices.\footnote{If the obligation amounts were not equal, B would have to pay whatever is left after set-off in the appropriate currencies.} Furthermore, the transfers of funds from A to C and from D to E will take place in USDC and ATOM, respectively, again without any interaction or need for a currency exchange service.

\begin{figure}[H]
\centering
\includegraphics[width=13cm]{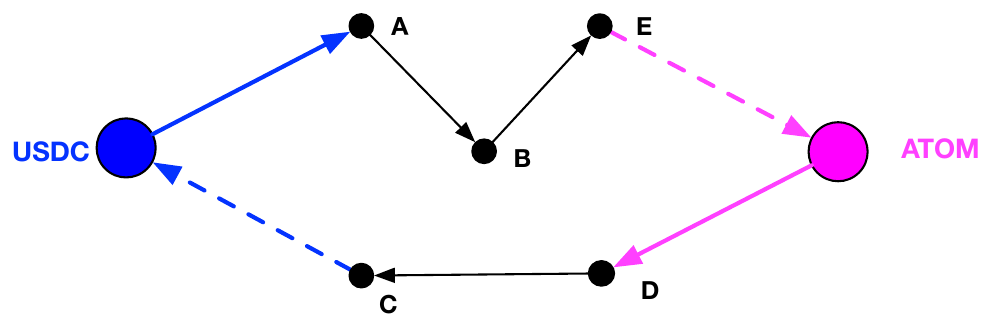}
\caption{\small\textbf{Two liquidity sources in the same cycle.} The use of multiple currencies can greatly improve the ability to discharge debt in the network.}
\label{fig:liquidity_couplediquidity}
\end{figure}

Now consider the case of Fig.\ \ref{fig:liquidity_couplediquidity}, where there are no complete payment cycles in a single currency and the symmetry of Fig.\ \ref{fig:liquidity_uncouplediquidity} is broken. Even this case can be solved via a \textit{single} cycle across \textit{two} liquidity sources. Once again, A's USDC can be used to pay C, and D's ATOM can be used to pay E. The currency transfers operate independently of each other, without an exchange service, according to the settlement records that define the settlement flow. 

\vspace{-0.4cm}

Our design thus enables a large number of currencies and liquidity sources to be utilized in the collaborative discharge of debt, greatly reducing working capital needs and various interest, exchange, and transfer fees. It opens profound new possibilities for different currencies and assets to be used in real world payments: so long as a small number of people are willing to use a currency, a much larger group stands to benefit. It promotes an open platform for participation of diverse actors and liquidity sources, and encourages development of new currency and credit protocols within a larger common framework for collective debt discharge, which yields numerous benefits.

\section{Cycles Protocol}

We now turn from payment system design to protocol description. We begin with a problem statement and a description of some user flows. We then describe the base privacy and settlement architecture in more detail before turning to discussion on design choices and extensibility.

\subsection{Problem Statement}
\label{problem}

The Cycles design requires execution of atomic multilateral settlement operations in a privacy-preserving and extensible credit environment across multi-scale graphs. We break down the problem statement into these parts.

\textbf{Atomic multilateral settlement} requires that settlement operations can occur “simultaneously,” across a large number of participants, in an all-or-none fashion, as a low-cost collective agreement to participate in cycles. These cycles can include obligations and acceptances across liquidity sources. This means, at a minimum, that all set-offs and balance changes in a single cycle must be executed atomically, across a set of parties who don’t necessarily all know each other (they only know their direct counterparties). Each party in a settlement cycle must be able to provide cryptographic proof for legal purposes that they are included in the cycle and that all operations within the cycle were executed atomically. More specifically, Alice must be able to prove that if her debtor receives a set-off notice that claims they don't have to pay her, then she also received one, meaning her debt to someone else was paid off or she received currency she accepts for payment.

\textbf{Privacy} requires first and foremost that obligation graphs are never revealed. This enables participants to submit their obligations without concern that any of their counterparty relationships will become known to third parties. However, since Cycles must perform a graph flow optimization algorithm (consisting of, at a minimum, addition and comparison operations), it must be able to execute this algorithm over a private graph. Secondly, privacy requires that obligation amounts are never revealed to third parties. This has limitations in that many existing asset ledgers do not have this property, and so it cannot be strictly enforced when assets enter and exit the Cycles system. 

\textbf{Extensible Credit Environment} requires
 the seamless integration of various liquidity sources, comprising existing assets and credit sources, alongside newly introduced ones. It should also accommodate user-specified and protocol-specified rules dictating their behaviour, such as issuance, pricing, interest, repayment, liquidation, etc. This entails integrating existing assets and credit sources for payments while facilitating the extension of new credit lines and the creation of novel credit assets.

\subsection{User Flows}

\textbf{Assignment Tenders.} The most basic use case is paying a bill. Like on any other blockchain, a user Alice can maintain a stablecoin balance on Cycles and use it to settle her debts -- whether to a supplier, service provider, creditor, etc. The major difference with Cycles is that Alice doesn't simply publish a transaction to send \$10 to Bob; she first declares that she owes Bob \$10. This illustrates the difference between an obligation and a tender. Almost any payment already involves an obligation; they just aren't represented on-chain.

Merely tendering \$10 of stablecoin after declaring you owe \$10 isn't particularly interesting. The magic comes from the network effect. Perhaps Alice's counterparty Bob doesn't accept stablecoins; he only accepts ATOM. However, he may owe Carol \$10, and Carol does accept the stablecoin. By having Alice, Bob, and Carol all declare their intents, Cycles can transfer Alice's stablecoin directly to Carol (without them being aware of each other) and publish set-off notices for everyone.

This use case can be expanded by adding more users and currencies. This is what it means to construct a graph. Across thousands of users, the network structure of the graph amplifies the benefits, conserving liquidity and reducing risk for everyone as more cycles form among users and liquidity sources. Users declare their obligations in a common unit of account\footnote{We leave multiple units of account for future work. In the end, Cycles must support graphs of different scales and different units of account. This is a ripe area for research and new directions on the idea of ``optimal currency areas''.} and publish their tenders and acceptances for any currencies supported by the network. At regular intervals (e.g.\ daily, monthly), solvers execute and find solutions that clear the most obligations for the most people with the least amount of liquidity, based on the published intents. 

\textbf{Overdraft Tenders.} Cycles also supports liquidity via overdraft tenders -- essentially, lending protocols implemented as smart contracts on the Cycles chain. These applications can directly integrate into the shared obligation graph. Users can access only the minimum amount of credit needed to optimize their payments, thus reducing interest costs. 

Consider a user without any stablecoin, but with some ATOM on their balance sheet. They could put up their ATOM as collateral in a lending protocol, allowing them to draw a stablecoin loan. However, on Cycles, they can structure this loan as an overdraft facility, submitting tenders against it, even before they draw it. The facility defines a set of rules for issuing repayment acceptances based on the collateral. The graph optimization will draw only the minimum amount of stablecoin credit needed from these acceptances to clear the debt. Further, the debt can be repaid automatically in the future in the form of on-chain obligations back to the overdraft facility. If Alice draws on her stablecoin overdraft to pay an obligation and later Bob owes her, solvers can optimize so that Bob's assets are effectively used to pay back Alice's overdraft, without any extra effort from Alice or Bob. This helps automate and improve working capital management, bringing increased efficiency and savings and better access to capital, and making DeFi more useful for real world use cases.

Cycles thus provides a more convenient and optimized way to use cryptocurrencies and lending facilities to support working capital concerns, reduce risk, and increase the likelihood of repayment. This opens the door to larger markets, more borrowers, reduced risk for lenders, and overall greater efficacy.

\textbf{Issuance.} In addition to being a platform for the development of diverse protocols for lending existing assets, Cycles also serves as a platform for protocols that issue new assets. In Section~\ref{overdraft} we called this \textit{overdraft with issuance}. It is a type of overdraft where the assets being lent do not already exist, but are created as part of the operation of lending. This includes collateralized stablecoin issuance protocols \cite{moneygod} as well as alternative credit protocols like mutual credit \cite{Litteraetal2017,DiniKioupkiolis2019} and trust networks \cite{circles}. By integrating directly into the graph, these protocols can benefit from the improved credit environments and new mechanisms for stability, fairness, and yield.

\textbf{External DeFi Integrations.} So far we've been discussing assets and smart contracts living on the Cycles chain, but Cycles is designed to be natively interchain. Lending protocols on external chains can also participate as overdraft facilities in Cycles through a system of ``virtual" tenders. To understand this, note that overdraft can always be turned into assignment by first drawing on the overdraft facility to pay yourself (i.e.\ a loan), and then tendering from assignment. The advantage of overdraft is that by only drawing to make payments, you draw the minimal amount you need when you need it, and don't pay interest fees for cash to sit on your balance sheet.

This is relevant for external lending protocols because it's always possible to borrow from such a protocol, and then move assets to Cycles, to then tender them from assignment. But this is inconvenient and expensive -- it takes multiple steps and you have to pay interest fees on the full amount you take out, even if you'll only end up needing less. Cycles can automate all of this and minimize the additional interest costs incurred from drawing from external overdraft facilities (ideally to 0) through a careful combination of locking and interchain communication. This provides new ways for existing lending protocols to increase their adoption and connect to the world of working capital finance.   

With these user flows in mind, we now turn to details of the base protocol for privacy and settlement.

\begin{figure}
\centering
\includegraphics[width=5in]{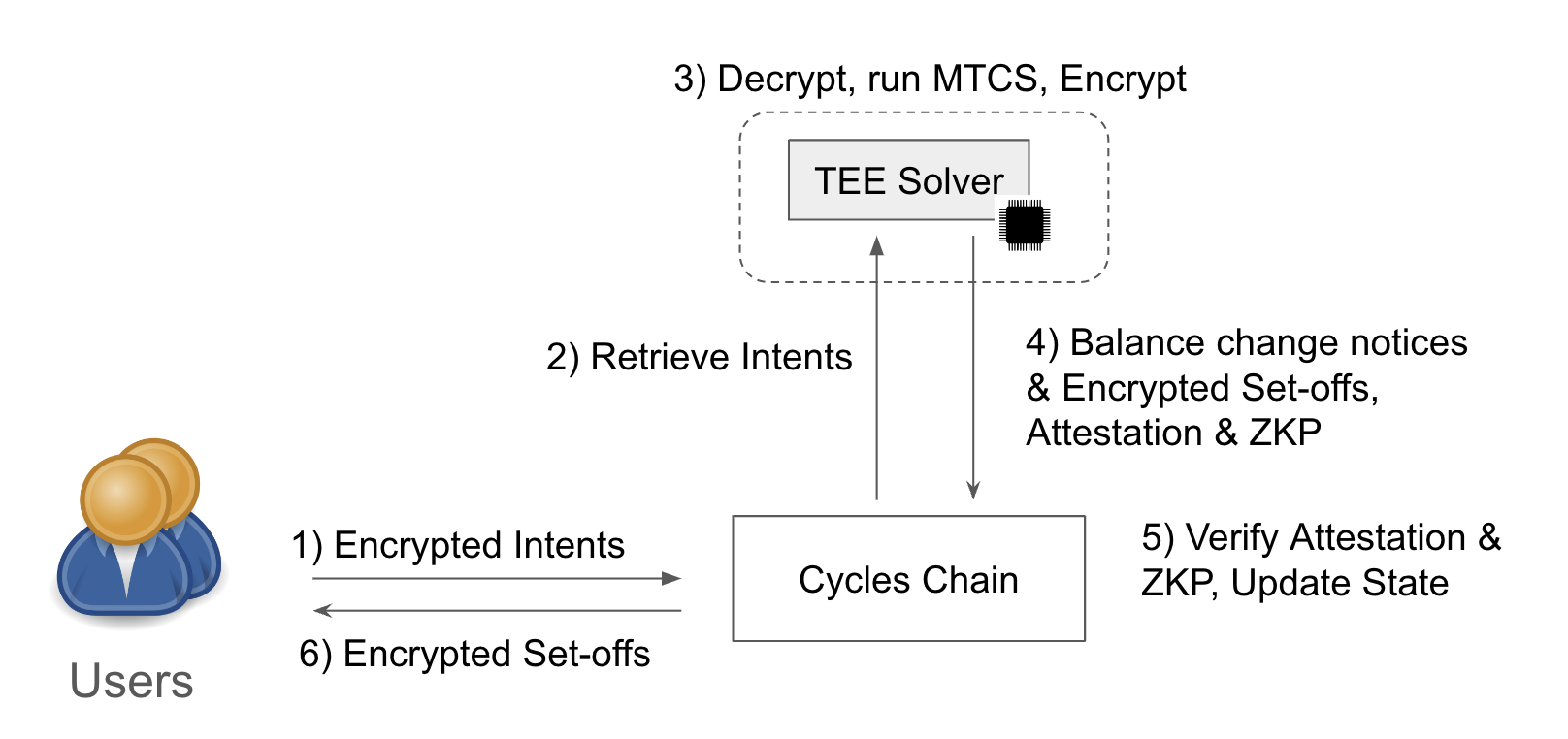}
\caption{\textbf{Cycles Architecture illustration}}
\label{fig:architecture}
\end{figure}

\subsection{Privacy \& Settlement Architecture} 
\label{privacy-architecutre}

Our goal is to develop a solution to the problem of Section~\ref{problem} that minimizes external institutional dependencies and maximizes the protocol's verifiability. That is, our goal is an Open Clearing Protocol. This is made possible by enabling technologies: using a combination of Byzantine Fault Tolerant (BFT) consensus \cite{buchman2016}, zero knowledge proofs (ZK)~\cite{hawk}, trusted execution environments (TEEs)~\cite{ekiden}, and obligation-based smart contract design~\cite{FleischmanDini2021}, a distributed system can carry out the atomic multilateral settlement operations in a private and extensible credit environment across multi-scale graphs.

The privacy design for Cycles takes the approach of a ``ZK+TEE Side Car". 
In order to provide privacy, we make use of TEEs to carry out the graph computation, and in order to ensure integrity, we use ZK proofs to verify the solution. The architecture is illustrated in Figure~\ref{fig:architecture}.


At a high level, the basic data flow is as follows. Users encrypt their intents (obligations, tenders, acceptances) and submit them to the chain. Periodically, the TEE retrieves all encrypted intents and decrypts them within the secure enclave. MTCS is run within the enclave, producing settlement records as a solution. Some of the settlement records pertain to changes in digital asset balances (the result of executing tenders), while others apply only to obligations. The latter are then encrypted to the public keys of the relevant users. The TEE produces a cryptographic attestation to correct execution within the enclave, as well as a ZK proof of the correctness of the result. The outputs -- including balance change notices, encrypted set-offs, and ZK proof -- are then submitted back to the chain, triggering the appropriate balance and state changes on-chain and committing to the set-offs, all in a single atomic action. User clients can then check the chain for set-off notices encrypted to them, and apply the result to their local accounting system. With this architecture, all graph information is stored encrypted on-chain, such that only a TEE and appropriate users can have access. 

For a graph $G$, the role of the TEE is to produce a \emph{valid} and \emph{optimal} flow solution $F := \textsf{Solve}(G)$. A \textit{valid} flow is a balanced flow, i.e. for each node, the flow in and out of the node is equal. An \textit{optimal} flow is one that satisfies a graph flow optimization as outlined in Section ~\ref{mtcsalgo}. We use the max-flow min-cost algorithm of MTCS as a default, though other optimality criteria can be defined. Correct execution of MTCS produces a valid and optimal solution.

The TEE will normally produce a \textit{remote attestation}, that is, a cryptographic commitment to correct execution of a particular program (in our case, MTCS). Since we want to guarantee validity even if the TEE is corrupted, we use a ZK proof, in addition to the remote attestation. To keep costs down, initially we will only use the ZK proof to ensure the posted solution is valid, while optimality depends on the TEE. This is enough to ensure atomicity of the solution and integrity of the balance sheets, as we only need to verify that a proposed solution $F$ is balanced and consistent with the inputs.
A ZK proof of validity $\textsf{IsValidFlow}(G,F)$  can be cheaper than a proof of optimality, since checking validity can be achieved in linear time $O(|G|)$, while checking optimality by including the entire flow solver algorithm in the ZK circuit could be asymptotically $O(|G|^2)$.

\textbf{Notation.}
To set out some notation, we have that each user $u$ is responsible for uploading their own obligations, defining the subgraph $G[u,:]$ of all edges radiating out from $u$ to all other nodes.
These edges will be posted as encrypted values to the Cycles chain, denoted 
$\hat{G} := \{ \textsf{Enc}(\textsf{xPub}, G[u,v]) \}_{u,v}$, where $\textsf{xPub}$ is the public key of the TEE that values are encrypted to.

We assume the Cycles chain has access to the ledger $A$, mapping addresses to account balances.\footnote{For simplicity here we assume only a single asset type as a source of liquidity, using assignment, though we can easily generalize to many sources and to overdraft.} Account balances are represented as an obligation in the graph $G$ where the debtor is a special symbol $\cal{L}$ (i.e.  $u$'s balance is an obligation from $\cal{L}$ to  $u$). Note that we are considering assets ``public” for simplicity, since here we focus on privacy for the obligation graph $G$.\footnote{In the future we can make these balances private by using a shielded pool.} Finally, we assume that all users have an acceptance of infinite size to the liquidity source. The obligation graph $G$ thus consists of the aggregate total of all the obligations uploaded by users, as well as all tenders and acceptances to and from the liquidity source $\cal{L}$.

\textbf{Zero Knowledge Proofs for Integrity Guarantees.}
A ZK proof can be used to validate the output of the graph flow algorithm, ensuring that only balanced flows are applied to the graph, and all user intents are respected. 
An idealized pseudocode for the smart contract verifier and included ZK proof is shown in Figure~\ref{fig:zkp}. 
The Cycles chain operates on an encrypted and committed graph denoted $\hat{G}$, while the ZK proof contains the plaintext graph $G$ as the witness.
The resulting flow solution $F$ will be committed on-chain as well. Since we only want to allow each user to learn about the set-offs pertaining to them, we publish an encryption of the corresponding set-off notices $\hat{N}$, along with a ZK proof confirming that these are in correspondence with the solution $F$, and that $F$ is a subset of $G$. In this way, the ZK proof in conjunction with the Cycles chain provides an end-to-end security guarantee about atomic multilateral settlement, without depending on the TEE at all.

We next explain several implementation details based on this high level plan.

\begin{figure}
Notation
\begin{itemize}
\item{$Enc(k, m)$: encrypt the message $m$ to public key $k$.}
\item{$xPub$: the public key of the TEE.}
\item{$pub_u$: the public key of user $u$.}
\item{$\hat{G}$: the set of obligations ($G$) encrypted to $xPub$.}
\item{$\hat{N}$: the set of set-off notices ($N$), encrypted to user public keys.}
\item{$F$: a flow solution, the result of running MTCS. A subset of $G$.}
\end{itemize}

Proof
\begin{itemize}
\item{Input: $[\hat{G}, \hat{N}]$}
\item{Witness: $[G, F, \{pub_u\}_u]$}
\item{Verify:}
\begin{itemize}
    \item \textbf{Decryption}: $G$ is an opening of commitment $\hat{G}$.
    \item \textbf{Ascertainment}: Each obligation and acceptance in $G$ has a valid signature from its debtor, and each tender has a valid signature from its sender.
    \item \textbf{Subset Flow}: Each element $f$ in $F$ corresponds to an element $g$ in $G$ with the same debtor and creditor and with $0 < f.amount \le g.amount$.
    \item \textbf{Balanced Flow}: For each user $i$, the sum of amounts in $F$ for creditor $i$ equals the sum of amounts in $F$ for debtor $i$.
    \item \textbf{Encryption}: For each element $f$ in $F$ with debtor $i$ and creditor $j$, $\hat{N}$ includes $Enc(pub_i, f)$ and $Enc(pub_j, f)$.
\end{itemize}
\end{itemize}
\caption{\textbf{Idealized ZKP pseudocode.} The proof must verify that the set-off notices were encrypted correctly from a flow solution that is balanced and corresponds to a subset of obligations that were signed by their debtors and correctly decrypted from those published on-chain.}
\label{fig:zkp}
\end{figure}

\textbf{Intent Posting and Batch Clearing.}
While we allow obligations to be entered into the system at any time (i.e.\ the system provides a continuous service just like any blockchain), the clearing events happen only periodically (such as daily, weekly, monthly) in batches. This means we have to pause the continuous service during the time we run the batch, so it is no longer practical to cancel an obligation once the MTCS algorithm has begun, or to submit new obligations to the current epoch. Obligations can continue to be submitted, but they will only be eligible for the subsequent epoch. 
Additionally, we need to take opportunities to do proactive validation of user transactions prior to the batch execution. At the least, some real-time spam prevention mechanism is required, for instance the payment of fees.

\textbf{Client-side Proactive Proving.}
In order to keep costs down and enable validation of user transactions prior to batch execution, clients can proactively provide their own ZK proofs of the validity of their obligations, acceptances, and tenders. This way intents are checked ``as they arrive” throughout the period, and become ``pre-validated,” at least enough to satisfy resource usage (e.g.\ pay fees), which reduces the amount of work that has to be done at clearing time. 

For simplicity, we have assumed that there is only a single obligation $G[u,v]$ between two parties $u$ and $v$. For multiple obligations, such as multiple invoices, these should be aggregated into a single total. This aggregation step does not have to be performed by the TEE; instead it could be carried out by the debtor. Similarly, decomposing the result into specific invoices that are (perhaps partially) discharged can be computed locally given the overall solution flow.

\textbf{Decomposing Large Proofs.} Proving the entire solution in one go for a large graph may be prohibitive. At the graph level, the entire flow solution could be proven in steps by decomposing it into smaller flows, and then constructing an aggregate proof across them. A standard technique for breaking up large ZK proofs into a series of smaller ones is to use Merkle trees to pass the shared state from one step of computation to another \cite{costello2015geppetto}.

For example, given flows $F_1, F_2, \cdots F_k$, we would compute the sequence of residual graphs and intermediate net positions that remain after applying each flow $G \overset{F_1}{\longrightarrow}, G'_1 \overset{F_2}{\longrightarrow}\cdots \overset{F_k}{\longrightarrow} G'_k$, and commit each residual graph and intermediate net position in a Merkle tree. At each step $i$, we would include the Merkle root of the previous step, which becomes part of the statement. We would then add to the ZK proof a Merkle tree update witness. Since each Merkle root appears in the subsequent proof, they bridge between the separate ZKPs of the residual graphs. A final proof can be constructed which aggregates over all the intermediate proofs.

\textbf{Blockchain as Coordinator for TEEs.}
In order to compute the solving function over the encrypted graph, a TEE must have access to the corresponding private key $\textsf{xPriv}$.
In order to provide redundancy, we must be able to share the $\textsf{xPriv}$ among multiple TEEs that can be used as backups in case one crashes. 
This requires the use of a blockchain to coordinate this process. First, the blockchain is used for validating remote attestation of the TEE. This is a certificate chain that can be posted alongside the public key, and has a root of trust signed by the manufacturer. The remote attestation includes a hash of the program binary, called the enclave hash. Second, the blockchain can be used to track the valid enclave hash. Before $\textsf{xPriv}$ is shared with a new TEE, it should be ensured that the posted program binary corresponds to the one approved on-chain. Ideally, before updating an enclave hash on-chain, the relevant software should undergo appropriate security audits and it should be confirmed that the software build can be deterministically reproduced from the published repository.

\textbf{Private Assets via Shielded Pool.}
The sender of an intent is leaked because ordinary transactions use native cleartext tokens to pay fees. Users can arrange to separate the fee paying address from their balance carrying address (i.e.\ the address which is debtor for their obligations, and which holds their tendered balances), using the obligation graph itself for privacy to unlink the two addresses. However, the balance carrying address is also a cleartext address, and so, while it can be hidden in the obligation graph, any balance changes that result from the multilateral settlement will still be leaked. This can be handled by adding a multi-asset ``shielded pool'' to the graph \cite{anoma, penumbra}, allowing balances to be kept and updated in private. This would require additional computation from both the client and the TEE to produce the necessary ZK proofs. These proofs would be publicly processed by the validators, enabling balance changes to remain private. Only the privacy of the graph itself would continue to rely on the TEE.

\textbf{Mitigating the Privacy Limitations of TEEs.}

In order to minimize the TEE's degrees of freedom and attack surface area, and to mitigate against grinding (i.e.\ running the TEE repeatedly on variations of the input to attempt data extraction), we follow a protocol whereby each TEE node must have an interactive roundtrip with the blockchain. This roundtrip makes use of a light-client protocol to ensure the TEE has an up-to-date view of the chain, and can only run with the chain's explicit permission. This limits the ability of nodes to access any secret data while in an isolated sandbox.

\textbf{Encrypted Queries.}
An extension of Cycles is to support user-defined graph queries. These have to be authorized by individual users. This logic is readily implemented as a smart contract running on the Cycles chain. One use of this is to allow the creditor of an obligation to decrypt the intent posted on the blockchain by the debtor. Another is to allow the construction of advanced credit ratings based on historical information in the payment graph.
Note that processing these in real time raises stronger questions about side channels than a batch operation.

\subsection{Discussion}

Here we discuss some of the design choices in the Cycles protocols and directions for future work. We touch on privacy, extensibility, and economics.

\textbf{Privacy.} 
We require the ability to perform computations on private inputs and produce private outputs. ZK proofs are effective for generating private outputs without divulging inputs; however, they do not facilitate private computation on the inputs themselves. In other words, they can provide privacy from the verifier, but not from the prover. Initially, they were utilized in blockchains for shielded pools, enabling confidential asset transfers where each participant possesses the necessary inputs to generate their own proofs. 

In Cycles, the objective is to maintain privacy of the obligation graph during multilateral operations. Unlike the case of shielded pools, where each user has access to all required inputs, no single user in Cycles has knowledge of all inputs. Consequently, relying solely on ZK proofs is inadequate since the prover would need access to all data. Many existing multilateral ZK proof systems, such as ZK rollups, expect solver agents to have access to all clear text inputs.

To prevent solvers from needing access to the graph in clear text, we must be able to execute operations on encrypted data. Broadly speaking, there are two approaches to solving this problem: multi-party computation (MPC) and trusted execution environments (TEEs). Fully Homomorphic Encryption (FHE) is often considered a third means of executing on encrypted data, but without using either MPC or TEEs in the decryption process, FHE alone is insufficient since a single key can decrypt the results.

MPC is a way of computing over encrypted data that distributes the trust required for privacy over a set of N nodes, such that K of them can fail yet privacy would still be ensured. Typically K is set to N/3, in order to balance privacy and availability guarantees. There are two main limitations to MPC. The first is performance -- while MPC is increasingly practical for managing keys, it suffers from poor performance for more complex computations, with network bandwidth as a bottleneck especially for geographically distributed nodes. Even more concerning than performance is the fact that K nodes could collude to decrypt user inputs; such a breach of privacy policy might not be detected at all, let alone be provable. Fully homomorphic encryption (FHE) can be used to improve performance of MPC (by trading off network IO for compute), but does not address the collusion hazard. At present, the only known way to address the collusion hazard is to additionally use TEEs.

TEEs such as Intel SGX and AMD SEV are based on trusted hardware that supports process isolation, ensuring that even the operator of the hardware cannot tamper with or inspect the process while it runs. This introduces a different kind of trust, since the manufacturer becomes the root of trust, and could in principle access private data within the TEE, or could attest to incorrect executions. Additionally, today's TEEs have been prone to vulnerabilities that have been disclosed and patched over several iterations in the past years \cite{sgxfail}. However, TEEs are currently orders of magnitude more performant than MPC,  and many of the vulnerabilities can be mitigated by careful protocol design. There is further promise for TEEs that are ``secure-through-physics" \cite{bellemare}.

Thus we use TEEs and work around their limitations. Ideally, TEEs are used to the minimal amount necessary to enable Cycles to satisfy its design requirements. In particular, we will use the TEE to keep the graph private while the solver runs, and we will rely on a ZKP as a backup to the TEE so that even if the TEE fails (privacy is violated), atomic multilateral settlement is still guaranteed. We leave practical implementations of MTCS via MPC and/or homomorphic encryption for future work. As noted in Section~\ref{privacy-architecutre}, Cycles can also be extended to utilize shielded pools to preserve privacy of transferred amounts, though we leave this also for future work. 

For side channels, we are not considering memory access pattern leakage. We think that when a batch graph algorithm is used, this will have minimal impact. Investigating mitigations such as Oblivious RAM is more important for real-time actions, which would leak more fine-grained information through access patterns. For forward secrecy, the TEE's public key can be changed with each epoch so that if the data from one epoch does leak it does not compromise data from other epochs. 

\begin{figure}
\centering
\includegraphics[width=5in]{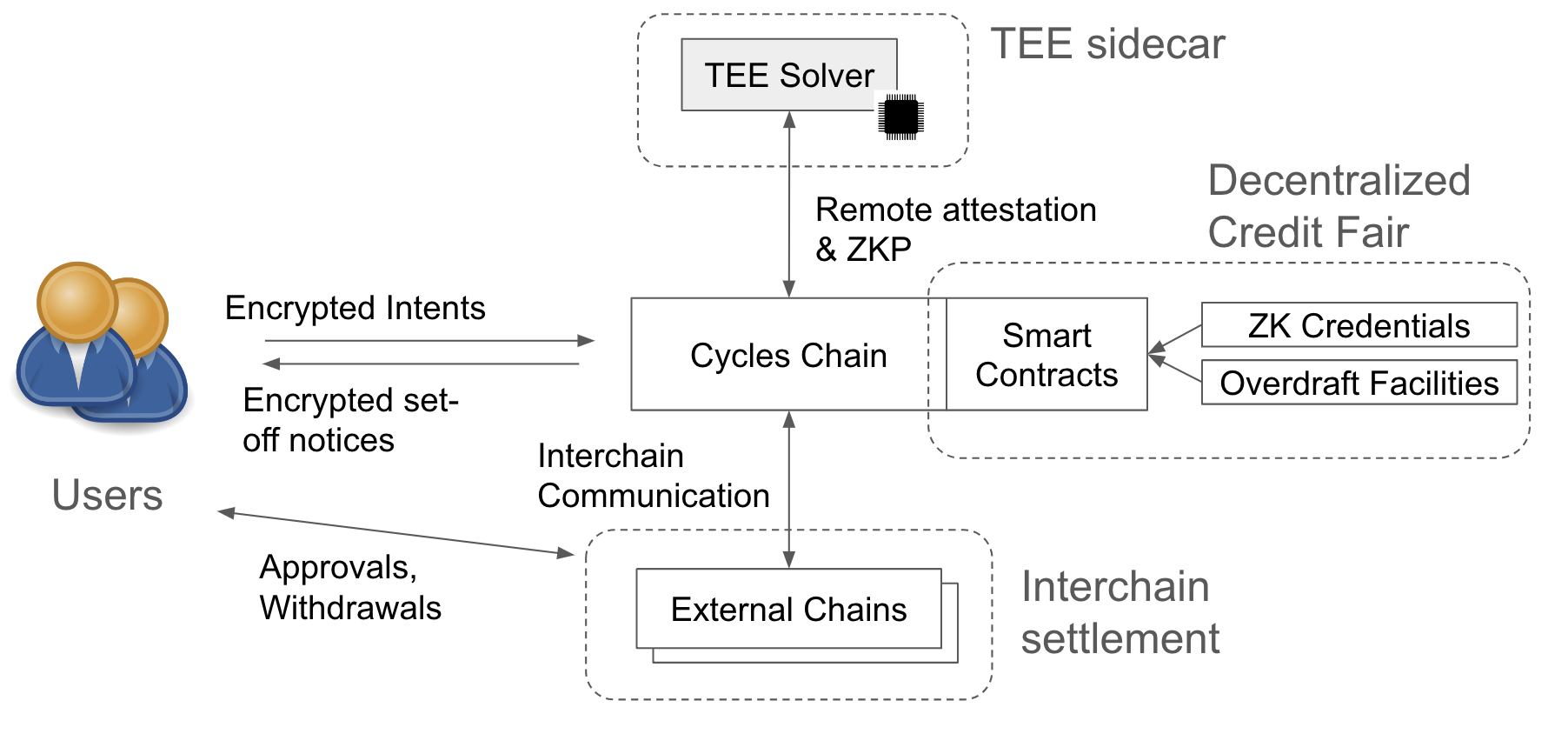}
\caption{\textbf{Extended Cycles Architecture illustration}}
\label{fig:architecture-extended}
\end{figure}

\textbf{Extensibility}

Cycles is built around an extensible, permissionless, Cosmos blockchain -- the root Cycles blockchain -- 
which serves as the single source of truth. It is a bulletin board for collecting (encrypted) user intents and set-off notices, a coordinator for the TEE-based computation, a verifier of the flow results, and an executor of the final settlements. It also serves as the coordination point for digital assets on remote chains, and as a host for new assets and credit instruments via a smart contract environment. This extended architecture is depicted in Figure~\ref{fig:architecture-extended}.

The Cosmos stack provides a rich set of tools for building reliable fault-tolerant and interoperable state machines. Cosmos was born out of a philosophy of sovereign and interoperable monetary zones~\cite{cosmos}, and has seen widespread adoption as a platform for launching blockchains and their token economies. Cycles builds on and extends the Cosmos technology and philosophy by providing a common language for describing payments, currencies, and credit protocols, and a new environment for decentralized finance (DeFi) to operate in, in the form of an open clearing club, that can more directly interface with the real world of trade and commerce.

To maximize impact, Cycles must support an extensible credit environment that can easily integrate with asset and credit facilities defined both externally and natively to the system. Assets and credit defined outside Cycles must be able to flow into Cycles, and new assets and credits must be able to be created within Cycles itself. Cosmos blockchains provide advanced capabilities for interoperability via the InterBlockchain Communication (IBC) protocol. IBC-enabled blockchains can easily transfer assets between one another in a trust-minimized way. More generally, the Interchain Accounts (ICA) protocol, built on IBC, allows one blockchain to perform arbitrary operations on another chain.

Cycles thus brings in liquidity from across the Cosmos ecosystem using IBC, allowing diverse assets to be used on the Cycles chain to discharge obligations, and making use of ICA to enable credit to be drawn from external Cosmos lending protocols in an automated and optimized way. Cycles is also designed to support assets and credit from Ethereum, which provides the largest source of permissionless lending protocols. Cycles thereby offers both the Cosmos and Ethereum DeFi ecosystems new opportunities for their credit facilitation to have more direct positive impact on real-world commerce and payments by integrating into the obligation graph. Cycles can also be connected to other blockchain environments either via custom bridges, or ultimately via widespread adoption of IBC. 

The Cycles chain uses a smart contract-based execution environment to support the creation of new assets and credit facilities natively on-chain. Smart contracts can themselves participate directly as firms in the graph, with their own obligations, tenders, and acceptances, and programmatic patterns for participation on their own account and on account of end-users. Every account is thus a node in the obligation graph, and every node in the obligation graph is an account. Native smart contract assets and overdraft facilities have the advantage of more direct composability with the graph flow optimization, allowing for lower costs and more functionality. These native facilities can also opt to use Cycles's native privacy architecture, having their state and execution logic managed by TEEs, using ZKPs as much as possible to guarantee integrity. Over time, the privacy architecture can be extended to further reduce the reliance on TEEs.

In Cycles, by default, every account is effectively a lending protocol by virtue of being able to offer repayment acceptances and in turn receive obligations. Repayment acceptances are one of the most fundamental design elements of Cycles. In a sense, Cycles is about making repayment acceptances (the general class of possible lending protocols) as accessible and flexible as possible. The primary use case for Cycles smart contracts, then, is building novel credit protocols that integrate with a larger obligation graph allowing them to take advantage of more sophisticated decision making. This includes better credit ratings, pricing and targeting of liquidity injection, managing non-performing loans, risk reduction, and ultimately new and more sustainable sources of yield. 

Overdraft facilities in Cycles can also define standing patterns of tenders and acceptances, as well as obligations, on behalf of users. Thus when a user opts into an overdraft facility, the facility can submit specific patterns of intents on the user's behalf. In this way Cycles can express diverse trust networks and credit protocols. Cycles can also incorporate exchanges, which can be understood in terms of tenders and acceptances between different currencies at different prices. Exchanges are thus a degenerate form of our model, devoid of obligations. Notably, decentralized exchange systems can be incorporated strategically into the graph in a manner that provides additional flow routes between liquidity sources. Referring back to Fig.~\ref{fig:liquidity_couplediquidity}, if there was no obligation from D to C, but instead we had an exchange between the two liquidity sources, A's USDC could be swapped to ATOM and paid to E, still completing the cycle. 

\textbf{Economics}

A core insight of Cycles is that extensibility in payment system design comes from interpreting all intents in terms of their underlying obligation structure. In particular, an acceptance is identical to an extension of credit, and becomes a future obligation: publishing an acceptance to a liquidity source is the same as saying you're willing to have that liquidity source owe you. This conceptualization allows us to see a certain equivalence between firms and liquidity sources -- firms are in themselves a kind of liquidity source (they can extend credit to others) and liquidity sources are a kind of firm (they can have their own behaviour and participate as agents in the graph). This opens profound possibilities for the design and integration of new credit protocols in payment systems. 

Cycles, then, is not just a protocol to clear your existing obligations with your existing assets, but rather a platform for accessing and issuing diverse sources of credit in an environment that is actually risk-reducing. Cycles, via MTCS, contributes to risk reduction across many dimensions, and thus enables a preferred environment for lending and borrowing. Individual enterprises reap the benefits of diminished risk due to regular balance sheet contraction, reducing leverage and supporting credit scores. By eradicating gridlocks within the payments graph using set-offs, MTCS reduces the payment risk, improves the days payable, and consequently reduces the late payment issue. This influence extends to loan repayments, as the system becomes a valuable tool for using circuit laws to recover non-performing loans (NPL) by extending new credit. Unlike central clearing houses, Cycles achieves this de-risking without introducing central counterparties to whom significant risk is transferred. Furthermore, Cycles presents an opportunity for more sophisticated credit risk assessment based on knowledge of the payment network. 

These risk reduction benefits apply in general, in the environment created by Cycles, across a diversity of lending and issuance protocols. Cycles can present itself to firms as a treasury dashboard with an overview of the firm's assets, overdraft facilities, and payment obligations. This creates an opportunity to make better-informed decisions about the use of available resources or to even automate the relevant tasks, to improve working capital conditions, reduce costs, and reduce risks. Such treasury functionality is normally reserved for large financial institutions, but via Cycles can be more readily made available to all.

Cycles is specialized around a batch operation that solves a flow optimization problem for accounts. The use of encryption and multilateral batching 
provides native censorship resistance, opens new economic design patterns, and allows specific economic objectives to be pursued during clearing. Cycles thus creates new economic possibilities for participants in credit networks. Leveraging the obligation graph allows participants to access better sources of credit, and thus save on interest fees and extractive terms. These benefits can in turn be shared with those that enable them across the graph. For each user, we can define a portfolio pricing rule, adding up their nominal assets minus their liabilities, and scaling by interest rates where appropriate. We can then incorporate this into the definition of our MTCS flow optimization to find sources of credit that minimize cost to the user.

We can thus conceptualize Cycles as a decentralized credit fair, where numerous participants gather to seek out the best sources of credit for themselves to unlock the greatest amount of economic activity with the least amount of cash liquidity. 

\section{Conclusion}

In this paper we introduced Cycles, an open clearing and issuance protocol. Cycles is based on a graph-theoretic understanding of the payment system that takes advantage of the network structure of credit relationships to clear the most debt with the least liquidity. This graph-based view enables new kinds of emergent economic behaviour by surfacing that firms themselves can serve as liquidity sources, and liquidity sources can behave like firms. The payment system design that results is unique in that it is based on atomic multilateral set-off operations which do not depend on a central counterparty or financial intermediary, and thus work to reduce both systemic and individual risk within the network. As a Cosmos blockchain, Cycles can preserve the privacy of its users and can be extended to enable diverse currencies and credit protocols to plug into the graph, allowing users to leverage their preferred assets for settlements and to access their most preferred sources of credit, while unlocking new opportunities for both costs and rewards to be shared across the network. Cycles thus not only innovates on critical issues in payments and financial systems architecture, addressing the liquidity pressure bearing down on small business around the world, it also offers a unique and compelling use case for blockchains, and provides a means for existing blockchain protocols to have more direct impact in the critical world of working capital concerns. Cycles: Respect the Graph.  

\section*{Acknowledgments}

This work builds directly on foundational prior work done with Giuseppe Littera \cite{Fleischmanetal2020}. The language developed in this paper was worked out over many sessions between the authors and other members of the Informal Systems Cycles team, especially Giuseppe Littera and Soares Chen. Other members of the Informal Systems team reviewed earlier drafts and provided valuable feedback, including Mark Yamashita, Juan Beccuti, Justin Wasser, and Arianne Flemming, as did Informal's investors, including personnel from CMCC Global, Maven11, Nascent, Cygni Capital, BKCM, and CMT Digital. Others who reviewed drafts and provided valuable feedback include Christopher Goes, Julio Linares, Maxime Monod, and Matthew Di Ferrante. We thank the reviewers from the Science of Blockchain Conference 2024 for their review on an earlier draft. We are grateful especially to Zarko Milosevic and Dan Elitzer for their incisive critiques and suggestions on an early version, which played a crucial role in refining this paper. The Interchain Foundation provided partial funding in 2022, which supported our research endeavors during the early phases of this work. Special appreciation is due to Informal Systems Inc, its members, and investors for providing the primary financial and organizational support for this work.

\setlength{\parskip}{0.8\baselineskip}
\bibliographystyle{plainurl}
\small
\addcontentsline{toc}{section}{References}
\bibliography{Bib/inmc_references,Bib/crypto}


\end{document}